\documentclass{iopart}
\usepackage{amssymb,amsfonts,amstext}
\usepackage{hyperref,epsfig}
\usepackage[svgnames]{xcolor}
\usepackage{pgf,tikz}

\renewcommand{\vec}[1]{\mathbf{#1}}
\newcommand{\arXiv}[1]{\href{http://arxiv.org/abs/#1}{\texttt{arXiv:#1}}}

\begin{document}

\review{Non-linear Dynamics and Primordial Curvature Perturbations from Preheating}
\author{Andrei V. Frolov}
\address{
  Department of Physics,
  Simon Fraser University\\
  8888 University Drive,
  Burnaby, BC Canada
  V5A 1S6
}
\ead{frolov@sfu.ca}
\date{April 13, 2010}

\begin{abstract}
  In this paper I review the theory and numerical simulations of non-linear
  dynamics of preheating, a stage of dynamical instability at the end of
  inflation during which the homogeneous inflaton explosively decays and deposits
  its energy into excitation of other matter fields. I focus on preheating in
  chaotic inflation models, which proceeds via broad parametric resonance. I
  describe a simple method to evaluate Floquet exponents, calculating
  stability diagrams of Mathieu and Lame equations describing development of
  instability in $m^2\phi^2$ and $\lambda\phi^4$ preheating models. I discuss
  basic numerical methods and issues, and present simulation results
  highlighting non-equilibrium transitions, topological defect formation,
  late-time universality, turbulent scaling and approach to thermalization.
  I explain how preheating can generate large-scale primordial (non-Gaussian)
  curvature fluctuations manifest in cosmic microwave background anisotropy
  and large scale structure, and discuss potentially observable signatures of
  preheating.
  
  \vspace{2em}
  \noindent\sl
  Dedicated to the memory of Lev Kofman, a friend and a mentor, without whom this article would never have been written.
\end{abstract}

\submitto{\CQG \hspace{-0.2em}(special cluster issue)}
\maketitle

\section{Introduction}

The idea of inflation (a period of rapid quasi-exponential expansion of the
Universe) neatly solves several long-standing issues in cosmology
\cite{Linde:2005ht, Linde:2005dd}, and has been spectacularly confirmed by
observations of the Cosmic Microwave Background (CMB) anisotropies
\cite{Komatsu:2010fb, Bennett:2010jb}. While the Universe is inflating, its
contents is cold. But eventually, inflation has to end and the field driving
the inflation must decay, depositing energy into high-energy particles. This
process, known as reheating, ``boils'' the vacuum and starts the thermal
history of the universe with the hot big bang. As universe continues to cool
down, it could undergo more phase transitions, which would happen at symmetry
breaking points of the theory. Very little is known about the fundamental
physics at these energy scales, and cosmological observations could be our
only source of information for the foreseeable future. No photons reach us
directly from this epoch as the universe is filled with hot plasma and is
opaque until recombination. Nevertheless, the expansion history of the early
universe is imprinted on the sky in the form of primordial curvature
fluctuations. With success of WMAP and the data from Planck soon to come, CMB
observations are reaching precision required to disentangle other subdominant
effects from Gaussian fluctuations due to simple inflation
\cite{Komatsu:2009kd}.

The most basic models of reheating involve inflaton decaying into one or more
other scalar fields. Among the most interesting are the ones where decay is
non-perturbative, for example proceeding through parametric resonance
naturally happening in chaotic inflation models \cite{Dolgov:1989us,
Traschen:1990sw, Kofman:1994rk, Shtanov:1994ce, Kofman:1995fi, Kofman:1997yn,
Greene:1997fu}, or tachyonic instability in hybrid inflation models
\cite{Linde:1993cn, GarciaBellido:1997wm, Felder:1998vq}. For all their
simplicity, these models have surprisingly rich physics involving
non-equilibrium phase transitions. While initial stages of preheating are
linear and instability development can be understood analytically
\cite{Kofman:1997yn, Greene:1997fu}, dynamics could be chaotic
\cite{Podolsky:2002qv}, and the field evolution quickly becomes inhomogeneous and
non-linear, so non-perturbative decay of the inflaton has to be studied
numerically \cite{Khlebnikov:1996mc, Prokopec:1996rr, Kasuya:1998td, Tkachev:1998dc, Felder:2000hj,
Felder:2001kt, Copeland:2002ku, GarciaBellido:2002aj, Podolsky:2005bw, Dufaux:2006ee,
Felder:2006cc}. In this paper I briefly review the theory of parametric
resonance, go over general results of numerical simulations of non-linear
field evolution during preheating, and discuss signatures of preheating that
could potentially be observed.

\section{Analytical Theory of Preheating}

\begin{figure}
  \begin{center}
  \begin{tikzpicture}[scale=0.8]
    \shade[top color=white,bottom color=LightSkyBlue] plot[id=R,domain=-2.34:2.34,samples=40] function{x**4/12};
    \draw[very thick] plot[id=V,domain=-3:3,samples=50] function{x**4/12};
    \draw[->] (-3,0) -- (3,0) node[right] {$\phi$};
    \draw[->] (0,-0.25) -- (0,7) node[above] {$V(\phi)$};
    \shade[ball color=red] (+2.535543701000,+5.034457448000) circle (.23) node[left] {\rotatebox{82}{slow roll}~~};
    \draw[->] (2.492616906,4.733082574) -- (2.353564078,3.850196815);
    \shade[ball color=red] (+0.000000000000,+0.250000000000) circle (.23);
    \node[above] at (0,0.5) {~oscillations};
    \node[above] at (0,2.5) {~~critical~~damping};
    \draw[->] (+0.2977500911,0.2506648756) -- (+0.7579414588,0.2805700982);
    \draw[->] (-0.2977500911,0.2506648756) -- (-0.7579414588,0.2805700982);
    \node[above] at (9,-1.1) {\epsfig{file=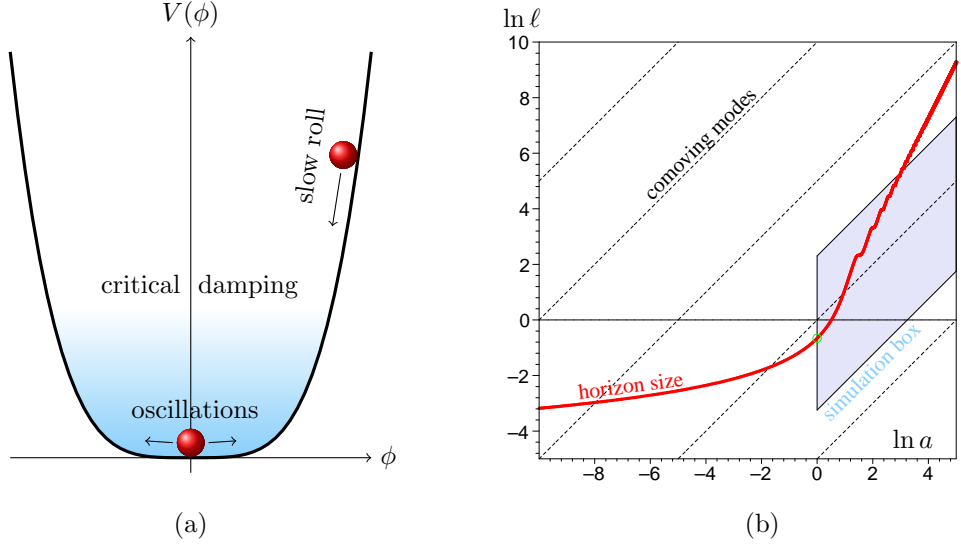, height=6.5cm}};
    \node[above] at (12,0) {$\ln a$};
    \node[above] at (5.5,7) {$\ln\ell$};
    \node[above] at (0,-1.5) {(a)};
    \node[above] at (9.5,-1.5) {(b)};
  \end{tikzpicture}
  \end{center}
  \caption{Dynamics of a single field chaotic inflation model with $\lambda\phi^4/4$ potential.}
  \label{fig:model}
\end{figure}

The basic model of reheating involves inflaton $\phi$ decaying into another
scalar field $\chi$. The action describing two interacting scalar fields
minimally coupled to gravity is
\begin{equation}\label{eq:action}
  S = \int \left\{ \frac{R}{16\pi G} - \frac{1}{2}\, g^{\mu\nu}(\phi_{,\mu}\phi_{,\nu}+\chi_{,\mu}\chi_{,\nu}) - V(\phi,\chi) \right\} \sqrt{-g}\, d^4 x,
\end{equation}
with potential $V(\phi,\chi)$ containing the terms responsible for field
masses and self-couplings, as well as their interaction. Polynomial field
operators up to fourth order are renormalizable, so one usually takes
$\frac{1}{2}\,m^2\phi^2$ or $\frac{1}{4}\,\lambda\phi^4$ inflaton potential
for chaotic inflation, and $\frac{1}{2}\, g^2\phi^2\chi^2$ coupling term
\cite{Kofman:1997yn, Greene:1997fu}. For simplicity, one can keep the decay
field $\chi$ massless. Couplings like $\frac{1}{2} \sigma\phi \chi^2$ are also
allowed and could be present \cite{Dufaux:2006ee}, although one would need a
$\chi^4$ self-interaction to keep the potential bounded from below. Models
with various combinations of these potential bits have been studied in the
literature; in this review, I will mainly focus on the one with quartic
potential \cite{Greene:1997fu}
\begin{equation}\label{eq:V:L4G22}
  V(\phi,\chi) = \frac{1}{4}\, \lambda\phi^4 + \frac{1}{2}\, g^2 \phi^2 \chi^2.
\end{equation}
In a flat homogeneous isotropic universe with Friedmann-Robertson-Walker metric
\begin{equation}\label{eq:frw}
  ds^2 = -dt^2 + a(t)^2 d\vec{x}^2,
\end{equation}
the field equations of motion are readily obtained from the action (\ref{eq:action}); they are
\begin{equation}\label{eq:phi}
  \ddot\phi + 3H\dot\phi + \left(-\frac{\Delta}{a^2} + \lambda\phi^2 + g^2\chi^2\right)\phi = 0,
\end{equation}
\begin{equation}\label{eq:chi}
  \ddot\chi + 3H\dot\chi + \left(-\frac{\Delta}{a^2} ~ \phantom{+ \lambda\phi^2} + g^2\phi^2\right)\chi = 0.
\end{equation}
Hubble parameter $H\equiv \dot{a}/a$ plays the role of friction term in field
dynamics. Its value is determined by the (average) total energy density
according to Friedmann equation
\begin{equation}\label{eq:H}
  H^2 = \frac{8\pi G}{3} \langle \rho \rangle,
\end{equation}
where the combined energy density of the two fields is
\begin{equation}\label{eq:rho}
  \rho = \frac{1}{2} \dot\phi^2 + \frac{1}{2} \dot\chi^2
       + \frac{1}{2} \frac{(\nabla\phi)^2}{a^2} 
       + \frac{1}{2} \frac{(\nabla\phi)^2}{a^2} + V(\phi,\chi).
\end{equation}
During chaotic inflation, potential energy of the inflaton causes Hubble
friction to be large, and the motion of the fields is over-damped. The
inflaton $\phi$ slowly rolls down the potential until the damping becomes
sub-critical, at which point it starts oscillating near the minimum of the
potential with decreasing amplitude, as illustrated in Figure~\ref{fig:model}a.
Evolution of the Hubble horizon size $L\equiv1/H$ and the physical wavelength
of comoving modes $\lambda \equiv 2\pi\, a/k$ is shown on Figure~\ref{fig:model}b.
During inflation $L$ changes slowly, so that slow roll parameter $\epsilon
\equiv \frac{\partial\ln L}{\partial\ln a} \ll 1$. When field is oscillating,
Hubble horizon size grows as $L\propto a^2$ according to the average equation
of state, which is $1/3$ for $\lambda\phi^4$ oscillator. Comoving modes stop
exiting and begin re-entering the horizon when $\epsilon=1$; this moment can
be taken as the end of inflation.

\begin{figure}
  \begin{center}
  \begin{tabular}{cc}
    \epsfig{file=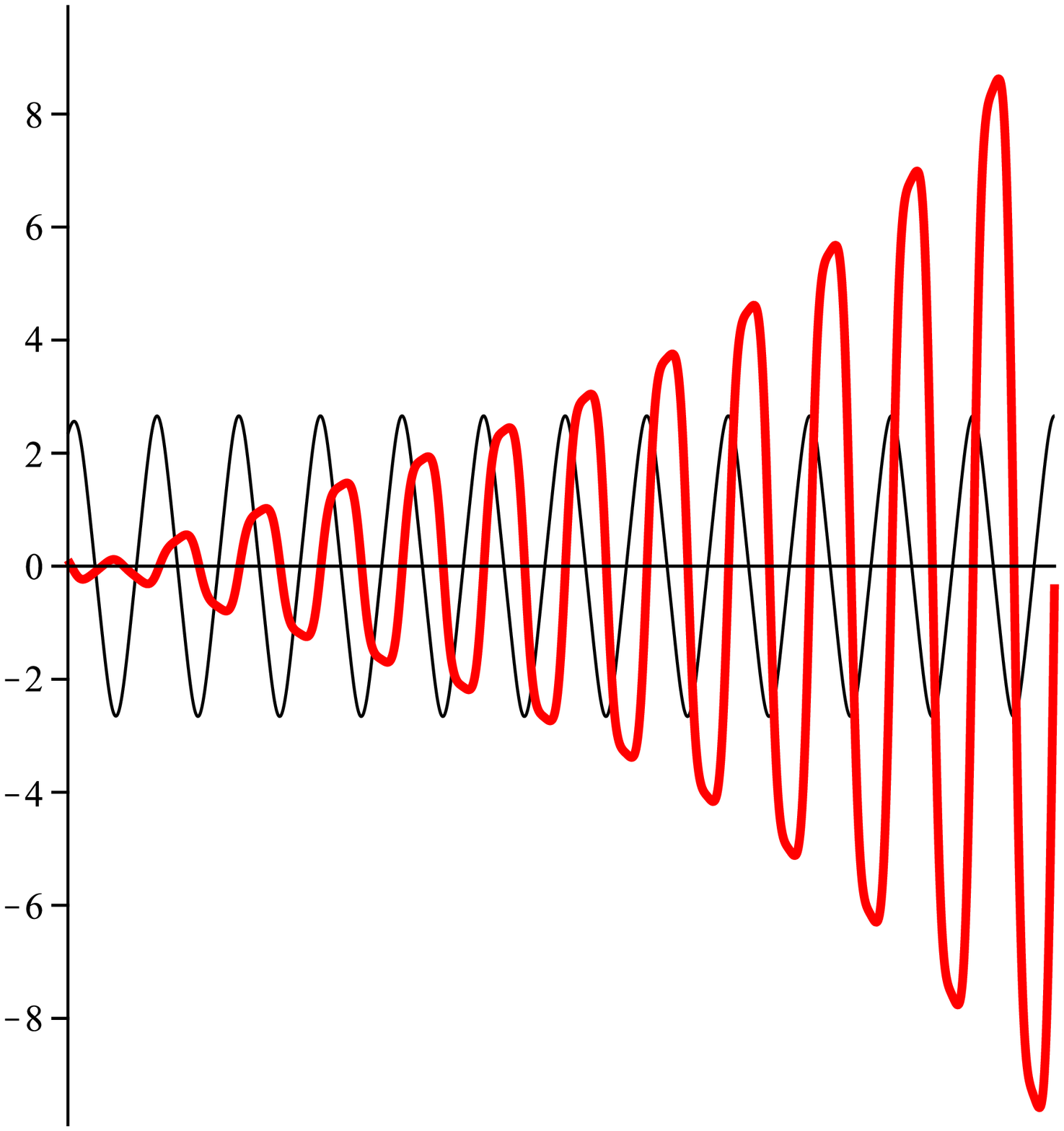, width=6cm} &
    \epsfig{file=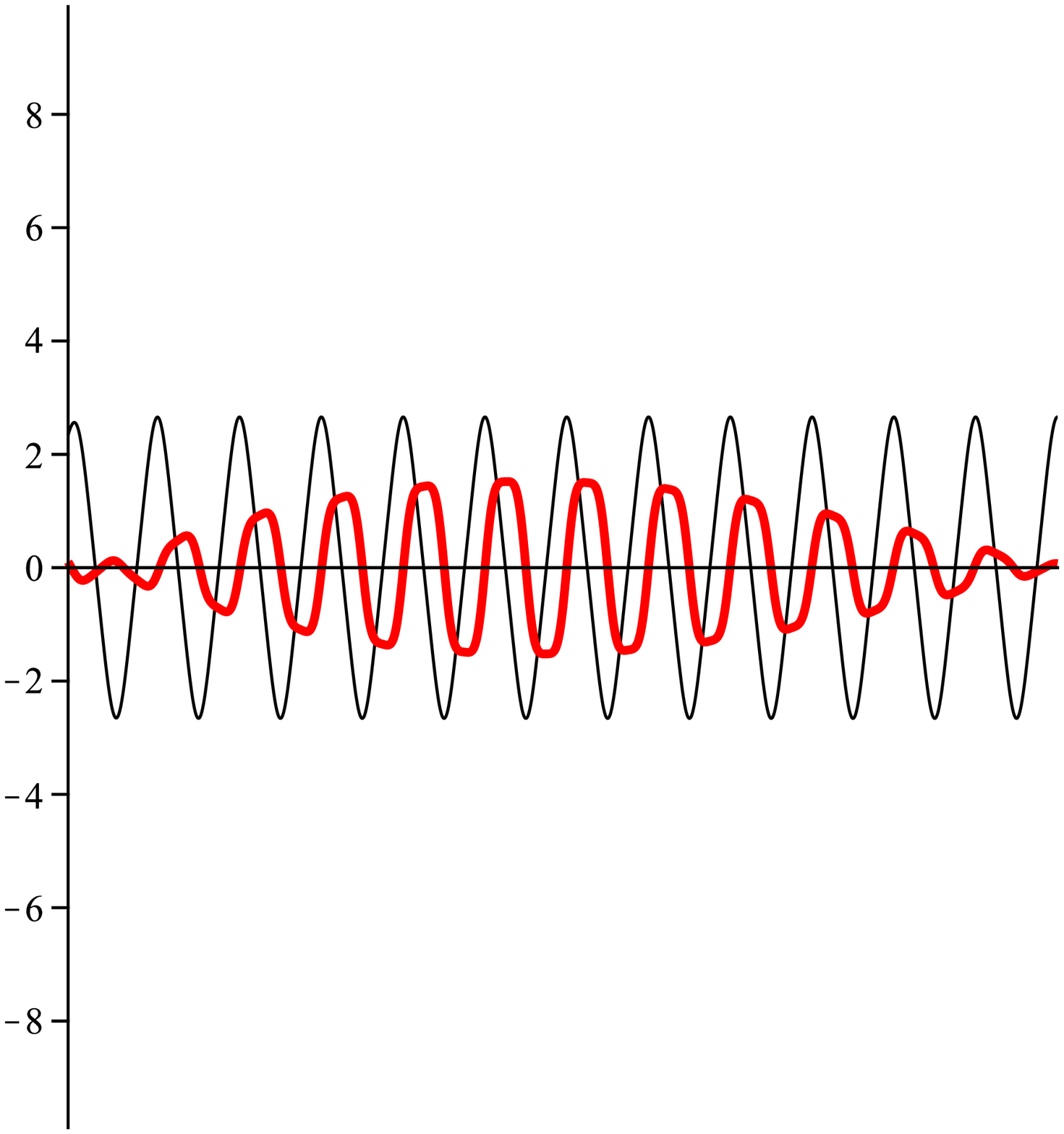, width=6cm} \\
    (a) $g^2/\lambda = 2.99$, $\kappa^2=0$ &
    (b) $g^2/\lambda = 3.01$, $\kappa^2=0$ \\
  \end{tabular}
  \end{center}
  \caption{Background oscillations of the inflaton $a\phi$ (black) and the zero mode of decay field $a\chi$ (red) for values of coupling slightly inside (a) and outside (b) of the first resonance band in Figure~\ref{fig:stab}b. Amplitude of $a\chi$ was scaled up by $3\cdot10^8$.}
  \label{fig:soln}
\end{figure}

Other fields coupled to inflaton feel its oscillations through modulation
of parameters in their equations of motion, such as the effective mass term
$g^2\phi^2$ in equation~(\ref{eq:chi}). Periodic modulation can lead to
parametric resonance and exponential growth of inhomogeneous excitations in
the fields coupled to inflaton. This is a fairly generic feature of chaotic
inflation models; let's see how this happens in our model~(\ref{eq:V:L4G22}).

It is very useful to scale variables so that the inflaton oscillations are
periodic and of constant amplitude. This is particularly easy in
model~(\ref{eq:V:L4G22}), which is conformally invariant apart from its coupling to
gravity. Switching to conformal time $d\eta \equiv dt/a$ and scaling the field
values according to their conformal weight, equation of motion for homogeneous
inflaton (\ref{eq:phi}) becomes simply $(a\phi)'' + \lambda(a\phi)^3 - a''\phi = 0$.
If one neglects the last term (which is small as $a\simeq\eta$), oscillating inflaton solution is
\begin{equation}\label{eq:bg}
  \phi(\eta) = \frac{\Phi_0}{a(\eta)}\, f(\tau),\hspace{1em}
  \tau \equiv \lambda^{\frac{1}{2}} \Phi_0 (\eta-\eta_0)
\end{equation}
where $\Phi_0$ is the amplitude of inflaton oscillations, and $f(\tau)$ is a
unit amplitude solution of the canonical anharmonic oscillator equation
$d^2f/d\tau^2 + f^3 = 0$, which can be written exactly in terms of Jacobi
elliptical cosine function, or its harmonic expansion \cite{Kiper:1984}
\begin{equation}\label{eq:cn}
  f(\tau) \equiv \text{cn}(\tau,2^{-\frac{1}{2}}) = 2^{\frac{1}{2}}\, \frac{4\pi}{T} \sum\limits_{n=1}^{\infty} \frac{\cos(n-\frac{1}{2})\frac{4\pi}{T}\tau}{\cosh(n-\frac{1}{2})\pi}.
\end{equation}
Function $f(\tau)$ is periodic with period
$T=\pi^{-\frac{1}{2}}\Gamma^2(\frac{1}{4})$, and its harmonic expansion is
exponentially converging, so only a few terms are needed to accurately
represent its shape. Substituting background inflaton solution (\ref{eq:bg})
into equation of motion (\ref{eq:chi}) for the coupled field $\chi$, and
rescaling variables the same way we did for inflaton, one obtains the
evolution equation for the Fourier mode of decay field $\chi_k$ with comoving
wavenumber $k$. In terms of rescaled parameters $\kappa \equiv
k/(\lambda^{\frac{1}{2}}\Phi_0)$ and $q \equiv g^2/\lambda$, it is
\begin{equation}\label{eq:lame}
  \frac{d^2 (a\chi_k)}{d\tau^2} + \left[\kappa^2 + q\, \text{cn}^2\left(\tau,2^{-1/2}\right)\right](a\chi_k) = 0.
\end{equation}
Exact solution of the oscillating inflaton $a\phi$ is shown in Figure~\ref{fig:soln},
along with homogeneous solution of the field $a\chi$ coupled to it, for two
slightly different values of the coupling $g^2/\lambda$. As you can see,
depending on the value of the coupling, evolution of the field $\chi$ can be
either exponentially unstable (\ref{fig:soln}a), or merely oscillatory
(\ref{fig:soln}b).

\begin{figure}
  \begin{center}
  \begin{tabular}{cc}
  (a) \hfill $\chi_k'' + \left[\kappa^2 + q \cos^2(\tau)\right]\chi_k = 0$~ &
  (b) \hfill $\chi_k'' + \left[\kappa^2 + q\, \text{cn}^2\left(\tau,2^{-1/2}\right)\right]\chi_k = 0$~~ \\
  \rotatebox{90}{\hspace{10em}$\kappa^2=k^2/(m^2 a^2)$}%
  \epsfig{file=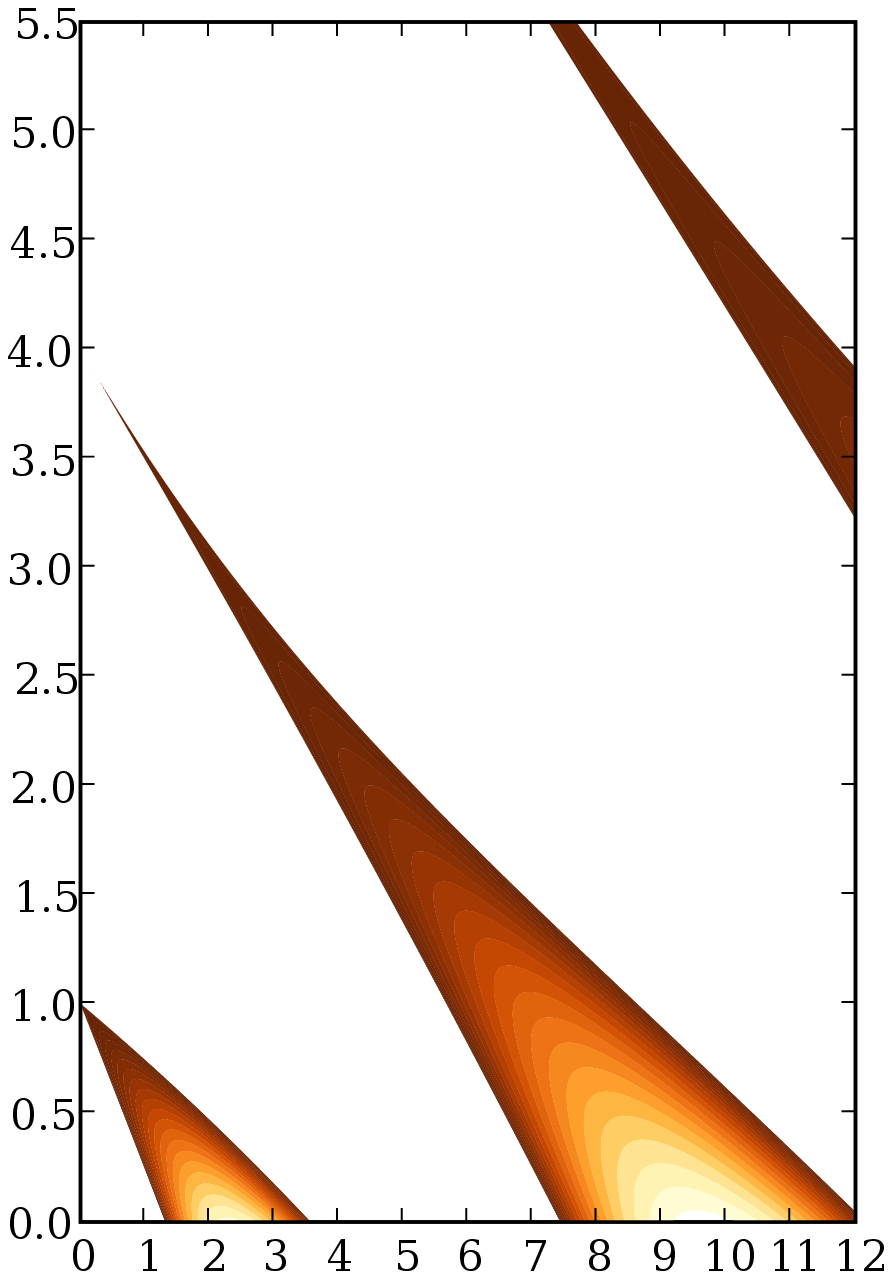, width=6cm} &
  \rotatebox{90}{\hspace{10em}$\kappa^2=k^2/(\lambda\Phi_0^2)$}%
  \epsfig{file=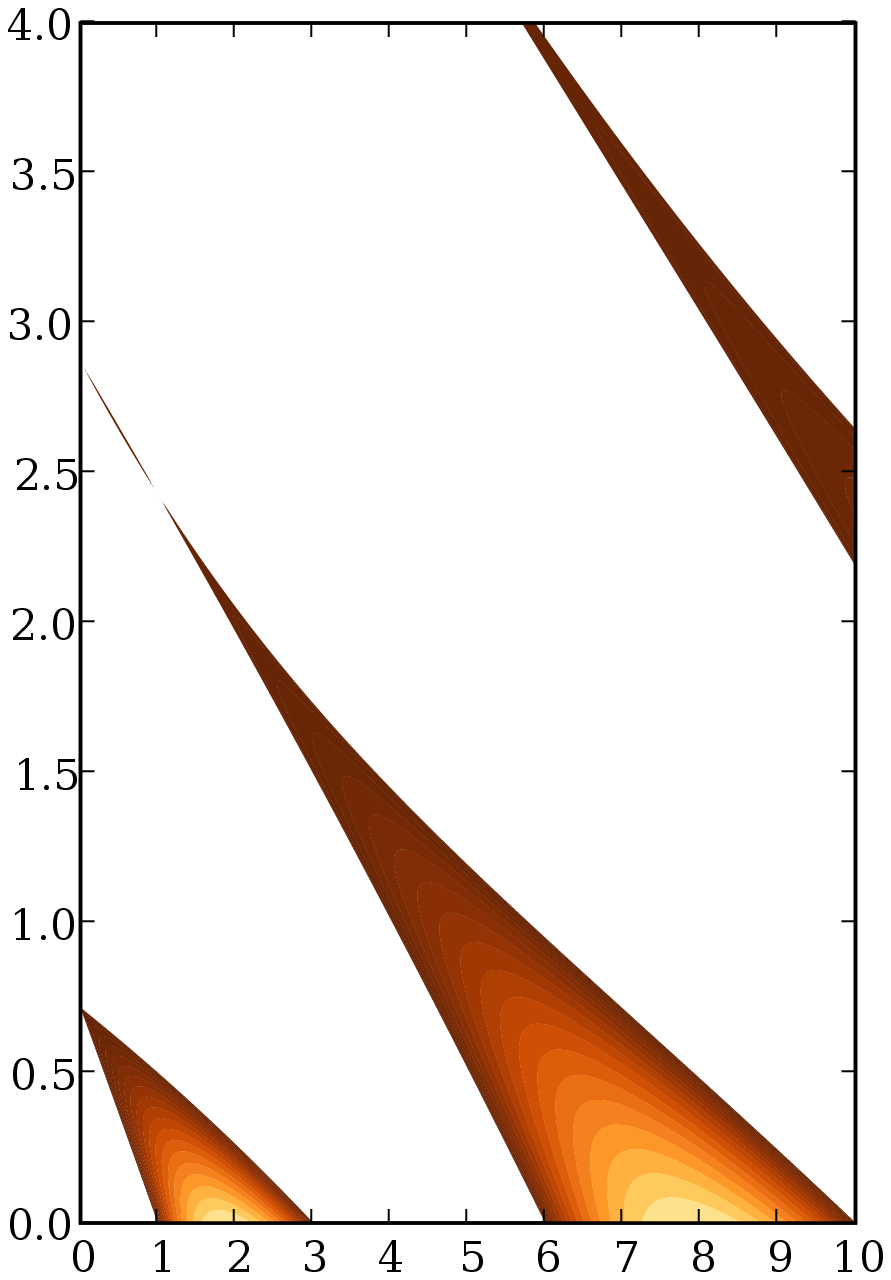, width=6cm} \vspace{-6pt}\\
  \hspace{2.2em} $q = g^2\Phi_0^2/(m^2 a^3)$ &
  \hspace{2.2em} $q = g^2/\lambda$ \\
  \end{tabular}
  \end{center}
  \caption{Stability diagram of Mathieu (a) and Lame (b) equations, using the same parametrization. White regions correspond to stable solutions, shaded regions are unstable, with brighter color corresponding to larger values of critical exponent $\mu$ (isolevels are spaced every $\Delta\mu = 0.01185$).}
  \label{fig:stab}
\end{figure}

Differential equations with periodic coefficients are studied in Floquet
theory, and are often encountered in other branches of physics as well (for
example, Bloch waves in condensed matter). Equation (\ref{eq:lame}), in
particular, is known as Lame equation, while its counterpart for harmonic
inflaton oscillations in $m^2\phi^2$ potential is Mathieu equation
\cite{Bateman:1955}. According to Floquet's theorem, equation (\ref{eq:lame})
admits solution of the form $e^{\mu \tau} P(\tau)$, where $\mu$ is a complex
number, and function $P(\tau)$ is periodic. Floquet exponent $\mu$ depends on
the parameters $\kappa^2$ and $q$, and can be calculated by explicitly
constructing such periodic function from the principal fundamental matrix
solution
\begin{equation}\label{eq:W}
  \mathbb{W}(\tau) = \left[\begin{array}{cc} \chi_1(\tau) & \chi_2(\tau) \\ \chi_1'(\tau) & \chi_2'(\tau) \\ \end{array}\right],
\end{equation}
made up from two independent solutions $\chi_1$ and $\chi_2$ with initial
conditions $\mathbb{W}(0) = \mathbb{I}$. Integrating principal fundamental
solution over a single period (which can be done efficiently and precisely
using numerical integration), and fixing coefficients in a linear combination
$c_1\chi_1(\tau) + c_2\chi_2(\tau) = e^{\mu \tau} P(\tau)$ to satisfy $P(T)=P(0)$ and
$P'(T)=P'(0)$, 
one finds that the value of Floquet exponent $\mu$ is given by
\begin{equation}\label{eq:mu:sqrt}
  e^{\mu T} = Q(T) + \sqrt{Q^2(T) - W(T)},
\end{equation}
where $Q$ and $W$ are two invariants of the matrix $\mathbb{W}$ under a similarity transformation
\begin{equation}\label{eq:Q}
  Q = \frac{1}{2} \tr \mathbb{W}, \hspace{1em}
  W = \det \mathbb{W} \equiv 1.
\end{equation}
The second invariant (Wronskian $W$) is conserved, so expression (\ref{eq:mu:sqrt}) simplifies to
\begin{equation}\label{eq:mu:cosh}
  \cosh \mu T = Q(T).
\end{equation}
Stability diagrams of Mathieu and Lame equations showing contour plots of
$\text{Re}\mu$ calculated in this fashion are presented on Figure~\ref{fig:stab}.
Wide bands in parameter space are unstable, with values of Floquet exponent
$\mu$ reaching as high as $0.237$ for Lame equation. If the value of coupling
lands you into one of these bands, inflaton will decay very efficiently into
$\chi_k$ particles for a wide range of wavenumbers $k$. This regime is known
as a broad parametric resonance.

Although not usually emphasized, the instability band structure of Lame and
Mathieu equations is quite similar, as elliptic cosine (\ref{eq:cn}) differs
from $\cos \tau$ by 4.3\% total harmonic distortion and 18\% larger period.
The real difference between $m^2\phi^2$ and $\lambda\phi^4$ models is how
parameters scale with expansion of the universe. While their values stay
constant for $\lambda\phi^4$ model (which is nearly conformally invariant),
for $m^2\phi^2$ model expansion drags the values of parameters toward the
stable point $\kappa^2=q=0$, eventually shutting off the resonance. Thus for
preheating to be efficient in $m^2\phi^2$ model, one needs a much larger
initial value of $q$.

\section{Tackling Non-linear Evolution: Numerical Methods and Issues}

\begin{figure}
  \epsfig{file=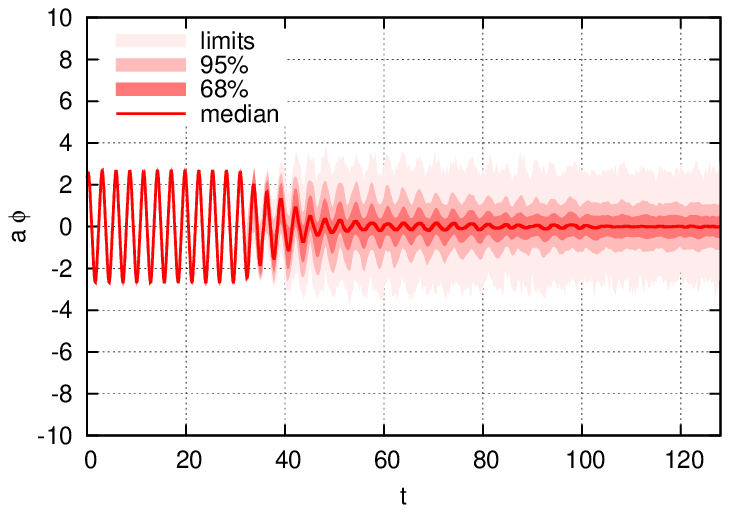, width=6.5cm}
  \hfill
  \epsfig{file=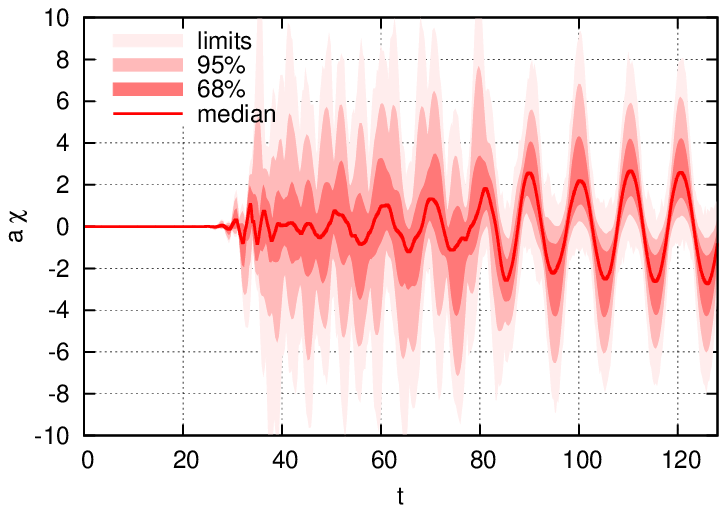, width=6.5cm}
  \caption{Evolution of distributions of inflaton (left) and decay product (right) in a full non-linear simulation with coupling at the maximal resonance $g^2/\lambda=1.875$.}
  \label{fig:decay}
\end{figure}

Broad parametric resonance amplifies quantum fluctuations of the fields,
creating real particles in a state far from thermal equilibrium. Instability
is exponentially rapid, and develops within a few dozen of inflaton
oscillation (as illustrated in Figure~\ref{fig:decay}), which is very fast on
cosmological time scales. Once the energy density of created particles becomes
comparable to that of the homogeneous inflaton, one can no longer treat the
evolution perturbatively, and non-linearity of the coupling and back reaction
of the created particles on the inflaton evolution has to be taken into
account. The most straightforward way to do it is to solve field evolution
equations numerically \cite{Khlebnikov:1996mc, Prokopec:1996rr}.

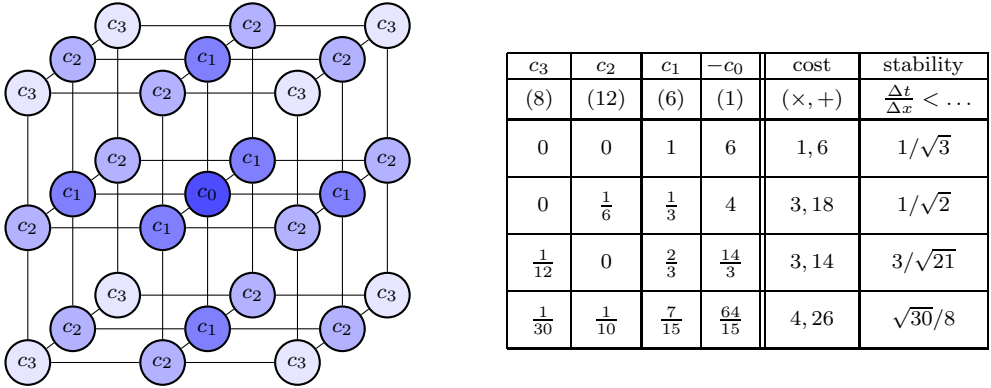
\begin{figure}
\begin{center}
\footnotesize
\begin{tikzpicture}[x=6em,y=6em,z={(2em,1.5em)}]
  % styles
  \tikzstyle{rank0}=[circle,draw=black,fill=blue!70,thick]
  \tikzstyle{rank1}=[circle,draw=black,fill=blue!50,thick]
  \tikzstyle{rank2}=[circle,draw=black,fill=blue!30,thick]
  \tikzstyle{rank3}=[circle,draw=black,fill=blue!10,thick]
  
  % nodes
  \node (000) at ( 0, 0, 0) [rank0] {$c_0$};
  \node (100) at ( 1, 0, 0) [rank1] {$c_1$};
  \node (010) at ( 0, 1, 0) [rank1] {$c_1$};
  \node (001) at ( 0, 0, 1) [rank1] {$c_1$};
  \node (I00) at (-1, 0, 0) [rank1] {$c_1$};
  \node (0I0) at ( 0,-1, 0) [rank1] {$c_1$};
  \node (00I) at ( 0, 0,-1) [rank1] {$c_1$};
  \node (110) at ( 1, 1, 0) [rank2] {$c_2$};
  \node (101) at ( 1, 0, 1) [rank2] {$c_2$};
  \node (011) at ( 0, 1, 1) [rank2] {$c_2$};
  \node (I10) at (-1, 1, 0) [rank2] {$c_2$};
  \node (I01) at (-1, 0, 1) [rank2] {$c_2$};
  \node (0I1) at ( 0,-1, 1) [rank2] {$c_2$};
  \node (1I0) at ( 1,-1, 0) [rank2] {$c_2$};
  \node (10I) at ( 1, 0,-1) [rank2] {$c_2$};
  \node (01I) at ( 0, 1,-1) [rank2] {$c_2$};
  \node (II0) at (-1,-1, 0) [rank2] {$c_2$};
  \node (I0I) at (-1, 0,-1) [rank2] {$c_2$};
  \node (0II) at ( 0,-1,-1) [rank2] {$c_2$};
  \node (111) at ( 1, 1, 1) [rank3] {$c_3$};
  \node (11I) at ( 1, 1,-1) [rank3] {$c_3$};
  \node (1I1) at ( 1,-1, 1) [rank3] {$c_3$};
  \node (1II) at ( 1,-1,-1) [rank3] {$c_3$};
  \node (I11) at (-1, 1, 1) [rank3] {$c_3$};
  \node (I1I) at (-1, 1,-1) [rank3] {$c_3$};
  \node (II1) at (-1,-1, 1) [rank3] {$c_3$};
  \node (III) at (-1,-1,-1) [rank3] {$c_3$};
  
  % edges
  \draw (III) to (II0) to (II1);
  \draw (I0I) to (I00) to (I01);
  \draw (I1I) to (I10) to (I11);
  \draw (0II) to (0I0) to (0I1);
  \draw (00I) to (000) to (001);
  \draw (01I) to (010) to (011);
  \draw (1II) to (1I0) to (1I1);
  \draw (10I) to (100) to (101);
  \draw (11I) to (110) to (111);
  \draw (III) to (I0I) to (I1I);
  \draw (II0) to (I00) to (I10);
  \draw (II1) to (I01) to (I11);
  \draw (0II) to (00I) to (01I);
  \draw (0I0) to (000) to (010);
  \draw (0I1) to (001) to (011);
  \draw (1II) to (10I) to (11I);
  \draw (1I0) to (100) to (110);
  \draw (1I1) to (101) to (111);
  \draw (III) to (0II) to (1II);
  \draw (I0I) to (00I) to (10I);
  \draw (I1I) to (01I) to (11I);
  \draw (II0) to (0I0) to (1I0);
  \draw (I00) to (000) to (100);
  \draw (I10) to (010) to (110);
  \draw (II1) to (0I1) to (1I1);
  \draw (I01) to (001) to (101);
  \draw (I11) to (011) to (111);
\end{tikzpicture}
\hfill
\footnotesize
\raisebox{8em}{
\begin{tabular}{|@{~~}c@{~~}|@{~~}c@{~~}|@{~~}c@{~~}|@{~~}c@{~~}||c|c|} \hline
    $c_3$ & $c_2$ & $c_1$ & \hspace{-0.8em} $-c_0$ & cost & stability \\ \hline
    $(8)$ & $(12)$ & $(6)$ & $(1)$ & $(\times,+)$ & \rule[-0.6em]{0pt}{1.7em} $\frac{\Delta t}{\Delta x} < \ldots$ \\ \hline\hline
    \rule[-1em]{0pt}{2.5em} $0$ & $0$ & $1$ & $6$ & $1,6$ & $1/\sqrt{3}$ \\ \hline
    \rule[-1em]{0pt}{2.5em} $0$ & $\frac{1}{6}$ & $\frac{1}{3}$ & $4$ & $3,18$ & $1/\sqrt{2}$ \\ \hline
    \rule[-1em]{0pt}{2.5em} $\frac{1}{12}$ & $0$ & $\frac{2}{3}$ & $\frac{14}{3}$ & $3,14$ & $3/\sqrt{21}$ \\ \hline
    \rule[-1em]{0pt}{2.5em} $\frac{1}{30}$ & $\frac{1}{10}$ & $\frac{7}{15}$ & $\frac{64}{15}$ & $4,26$ & $\sqrt{30}/8$ \\ \hline
\end{tabular}}
\end{center}
\vspace{-1em}
\caption{Three-dimensional spatial discretization stencil (left) and summary of coefficients (right) for minimal (top) and three isotropic discretization schemes.}
\label{fig:disc}
\end{figure}

Several codes are available for this purpose, most notably LATTICEEASY by Gary
Felder and Igor Tkachev \cite{Felder:2000hq, Felder:2007nz} which is widely
used and modified, my own DEFROST~\cite{Frolov:2008hy} which offers improved
performance and visualization capabilities, and a new GPU-accelerated CUDAEASY
by Jani Sainio \cite{Sainio:2009hm}. A pseudo-spectral code PSpectRE has
just been released as well~\cite{Easther:2010qz}. Most of the implementations
(with the exception of PSpectRE) opt for a finite difference method to solve
non-linear partial differential equations (\ref{eq:phi},\ref{eq:chi}).
The fields are discretized on a cubic spatial grid of spacing $dx$, and the
spatial differential operators are approximated by finite differences
\begin{equation}\label{eq:disc}
  \Delta X = \frac{D[X]}{(dx)^2}, \hspace{1em}
  (\vec{\nabla} X)^2 = \frac{G[X]}{(dx)^2}.
\end{equation}
Discretizations of Laplacian operator involving only 26 nearest neighbours of a point
\begin{equation}\label{eq:D}
  D[X] \equiv \underbrace{\sum\limits_{x-1}^{x+1}\sum\limits_{y-1}^{y+1}\sum\limits_{z-1}^{z+1}}\limits_\alpha
    c_{\text{d}(\alpha)} X_\alpha
\end{equation}
is going to be second order accurate, but truncation error can be made
isotropic to fourth order \cite{Patra:2005} by taking coefficients $c_\alpha$
as summarized in Figure~\ref{fig:disc}. A critical issue for accuracy of
long-term cosmological simulations is discretization of gradient terms in
energy density (\ref{eq:rho}). Discretized energy is not necessarily
conserved by discretized equations of motion, and gradient energy leaking off the
grid affects equation of state and leads to large cumulative errors in
expansion history of the universe \cite{Chambers:2007se,Bond:2009xx}. The best
way to avoid this pitfall is to discretize Lagrangian (\ref{eq:action})
directly. One can show that the proper discretization of gradient square, variation of
which leads to Laplacian discretization (\ref{eq:D}) in equations of motion, is
\begin{equation}\label{eq:G}
  G[X] \equiv \frac{1}{2} \underbrace{\sum\limits_{x-1}^{x+1}\sum\limits_{y-1}^{y+1}\sum\limits_{z-1}^{z+1}}\limits_\alpha c_{\text{d}(\alpha)} (X_\alpha-X_0)^2.
\end{equation}
Once the spatial operators are discretized, time evolution problem becomes a
system of coupled ordinary differential equations, which can be integrated
using any of the usual methods. DEFROST uses leapfrog scheme, which is simple,
fast, and second order accurate in time. Higher order scheme could be used
if needed. Symplectic integrator developed in \cite{Bond:2009xx} is capable of
reaching machine precision levels, or hybrid integrator of Huang
\textit{et.\ al.}\ \cite{Barnaby:2009wr} could be used if time
operator splitting is not possible.

\section{Non-Linear Dynamics, Thermalization and Universality within Horizon}

Non-linear dynamics that soon takes over the evolution of scalar fields can be
rather non-trivial. Preheating is essentially a non-equilibrium phase
transition, and a lot of interesting things can happen. Over the years,
detailed numerical studies have been carried out for many parametric resonance
\cite{Khlebnikov:1996mc, Prokopec:1996rr, Tkachev:1998dc, Podolsky:2005bw, Dufaux:2006ee,
Felder:2006cc} and tachyonic \cite{Felder:2000hj, Felder:2001kt, Copeland:2002ku,
GarciaBellido:2002aj} preheating models. In this section, I highlight some
features of non-linear dynamics in preheating on horizon scales. This is local
physics as far as cosmology is concerned, as the Hubble horizon size at the
end of chaotic inflation is tiny (roughly $1$m redshifted to the present day).

\begin{figure}
  \begin{center}
  \begin{tabular}{cc}
    \epsfig{file=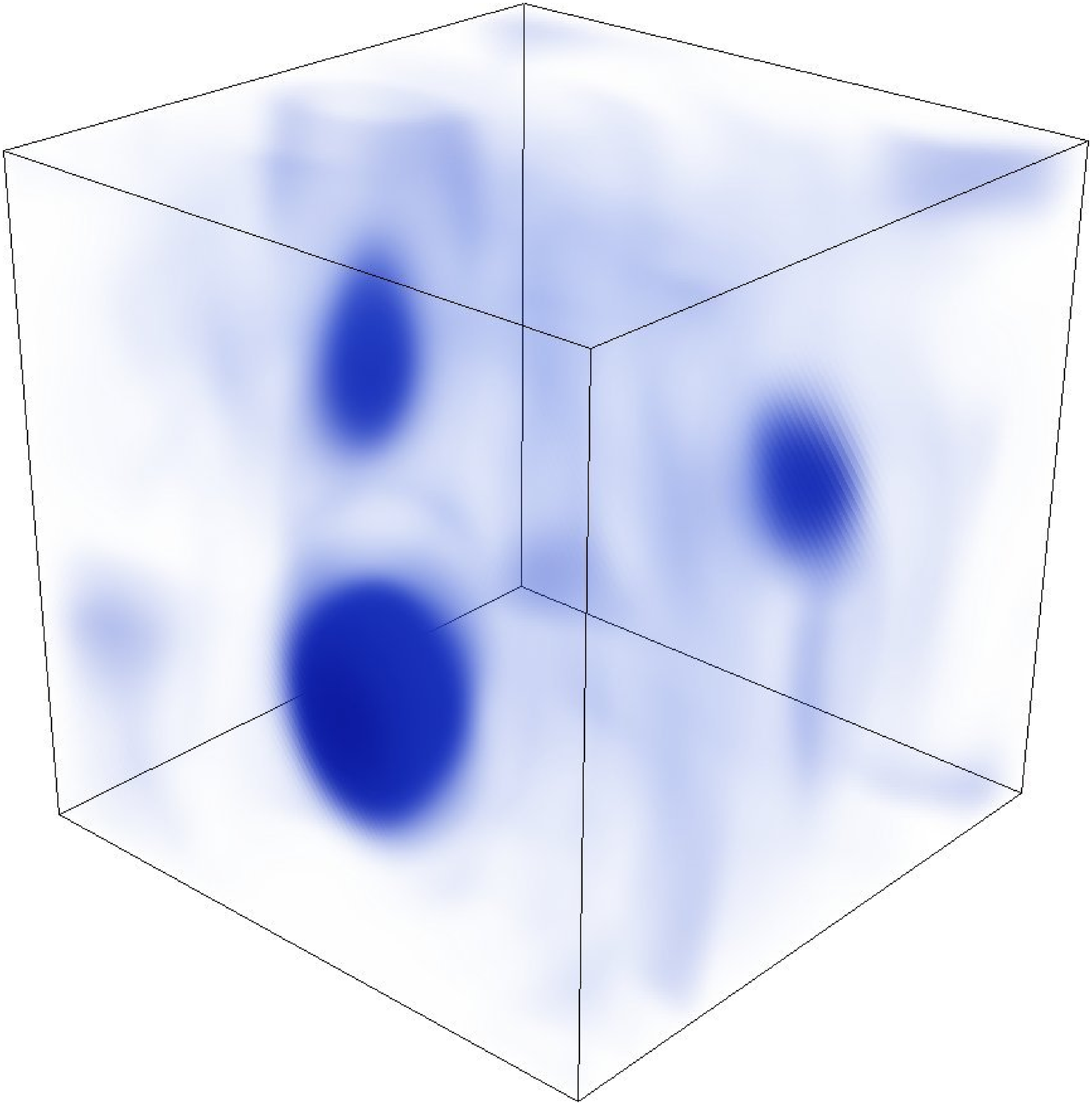, width=6cm} &
    \epsfig{file=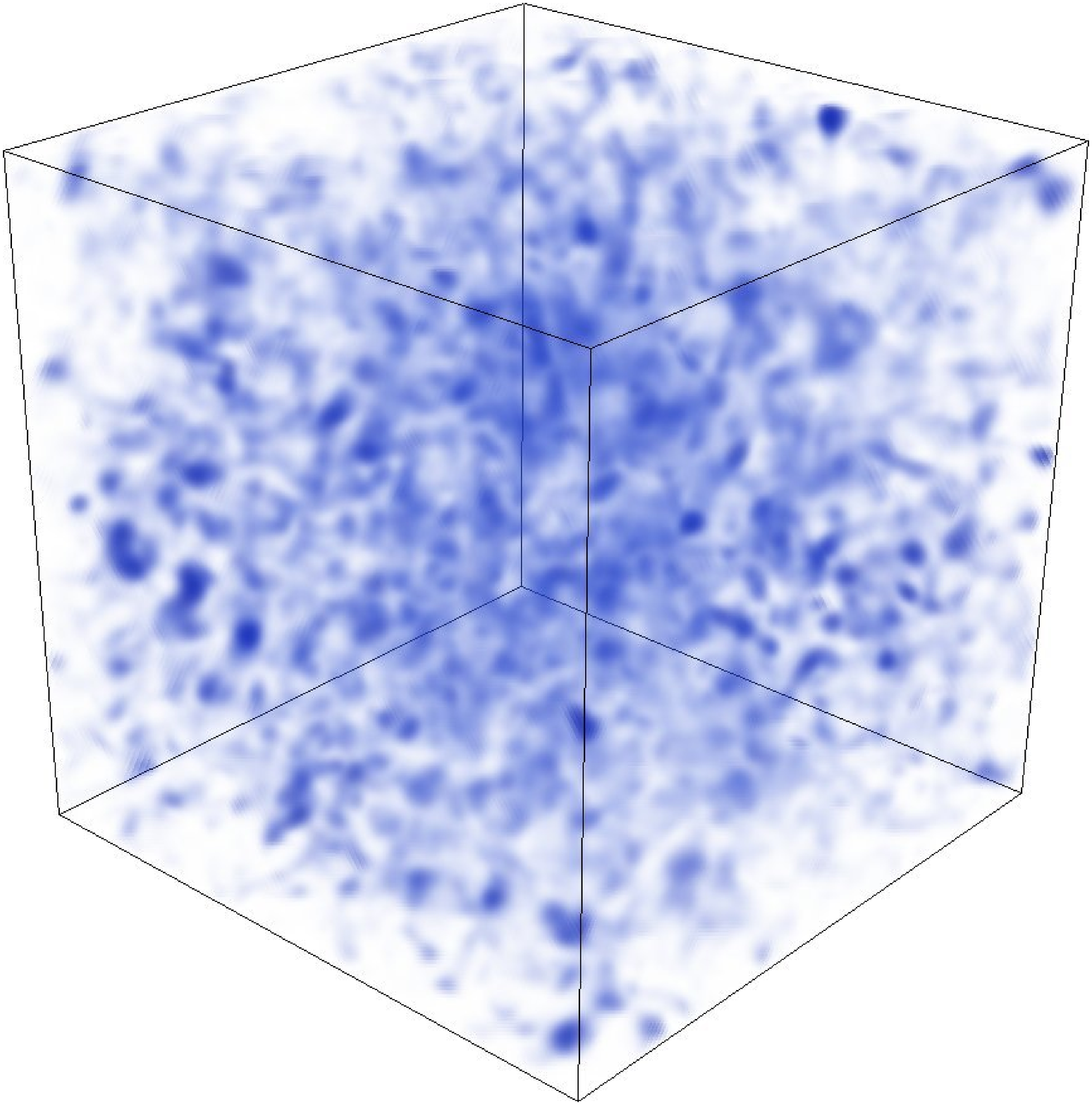, width=6cm} \\
    (a) $t=36.75$ &
    (b) $t=128.00$ \\
  \end{tabular}
  \end{center}
  \caption{Energy density $\rho$ inside the simulation box soon after onset of instability (a) and during subsequent evolution (b) in preheating model (\ref{eq:V:L4G22}).}
  \label{fig:rho}
\end{figure}
\begin{figure}
  \begin{center}
  \begin{tabular}{cc}
    \epsfig{file=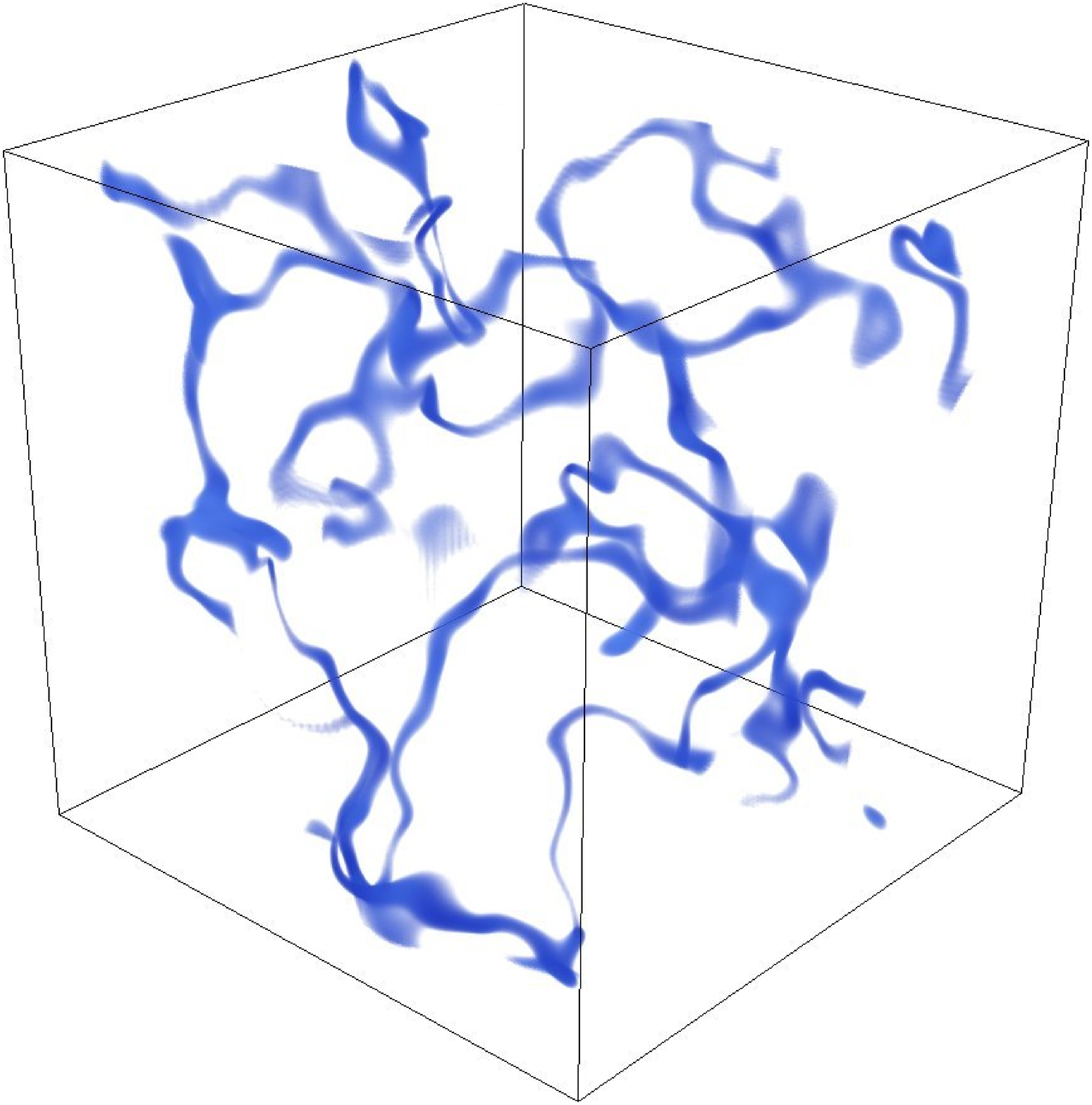, width=6cm} &
    \epsfig{file=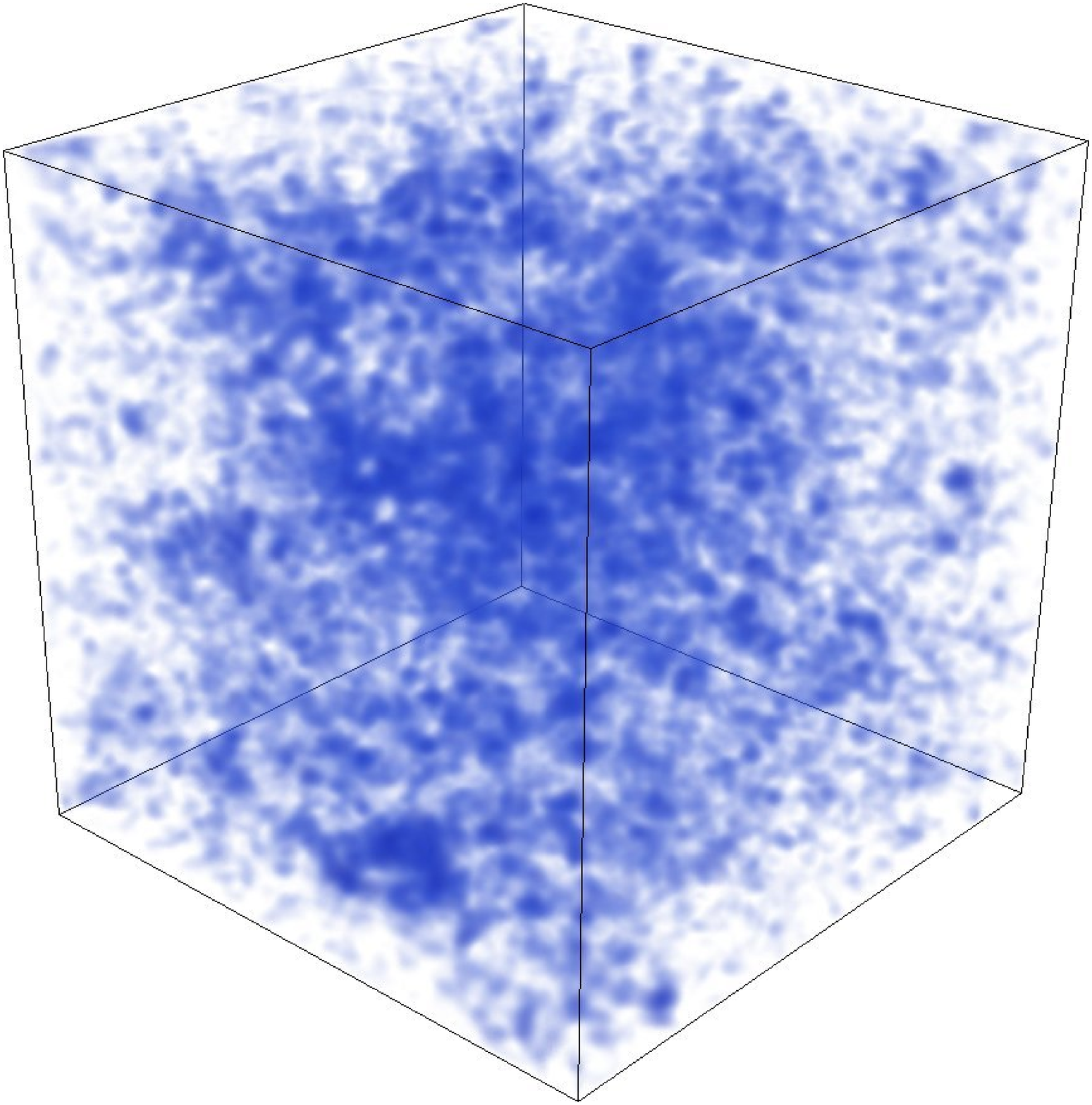, width=6cm} \\
    (a) $t=103.00$ &
    (b) $t=512.00$ \\
  \end{tabular}
  \end{center}
  \caption{Transient formation of topological defects (a) and late-time energy density configuration (b) in preheating model (\ref{eq:V:L4U1V2}) with global $O(2)$ symmetry.}
  \label{fig:strings}
\end{figure}

Quantum fluctuations that fall into unstable bands of Figure~\ref{fig:stab}
are amplified and become classical with large occupation numbers. Once linear
instability develops, the field configuration becomes very inhomogeneous and
particle description gets complicated by non-linear coupling. A useful tracer
of field dynamics is the evolution of the total energy density (\ref{eq:rho}),
which is an adiabatic invariant for rapidly oscillating fields. Evolution of
energy density $\rho$ in preheating model (\ref{eq:V:L4G22}) with
$g^2/\lambda=1.875$ is shown in Figure~\ref{fig:rho}. For that value of the
coupling, long wavelength fluctuations grow fastest, leading to formation of
large blobs as shown in Figure~\ref{fig:rho}a. Once density contrast increases
to about one, non-linear interaction kicks in and the density fragments to
much smaller scales as shown in Figure~\ref{fig:rho}b. In particle
description, this could be viewed as upscattering of modes to higher momenta
by non-linear interaction term.

Just as it happens in phase transitions, one can also produce topological defects
during preheating, if the theory allows them. For example, in preheating model
with $O(2)$-symmetric potential with a small field vacuum expectation value $v$
\begin{equation}\label{eq:V:L4U1V2}
  V(\phi,\chi) = \frac{1}{4}\, \lambda (\phi^2 + \chi^2 - v^2)^2,
\end{equation}
global cosmic strings can form \cite{Kasuya:1998td, Tkachev:1998dc}. Initial instability in
this model develops similarly to (\ref{eq:V:L4G22}) with $g^2/\lambda=2$, but
once the energy density dilutes enough to fill the ring at the bottom of
potential, cosmic strings are produced. String cores ($\phi^2+\chi^2 <
v^2/40$) soon after formation are shown in Figure~\ref{fig:strings}a. String
loops within horizon are transient, and eventually will collapse and
annihilate. After a while, energy density configuration once again reaches a
highly fragmented state shown in Figure~\ref{fig:strings}b.

An important open question is exactly how and when thermalization after
preheating happens. Characteristic momentum of particles produced during
linear stage of preheating is determined by the instability band structure.
Typically, it is significantly less than that of thermal equilibrium value, so
particles need to upscatter through non-linear interactions, and
thermalization can be delayed by a long time \cite{Kofman:1997yn}. This is
supported by numerical simulations, which show slow scaling regime in
evolution of field occupation numbers in $\lambda\phi^4$ model
\cite{Micha:2002ey,Micha:2004bv}.

\begin{figure}
  \begin{center}
  \begin{tabular}{rr}
    ~~(a)\hfill $V = \frac{1}{4}\, \lambda \phi^4 + \frac{1}{2}\, g^2 \phi^2 \chi^2$ &
    ~~(b)\hfill $V = \frac{1}{2}\, m^2 \phi^2 + \frac{1}{2}\, g^2 \phi^2 \chi^2$ \\
    \epsfig{file=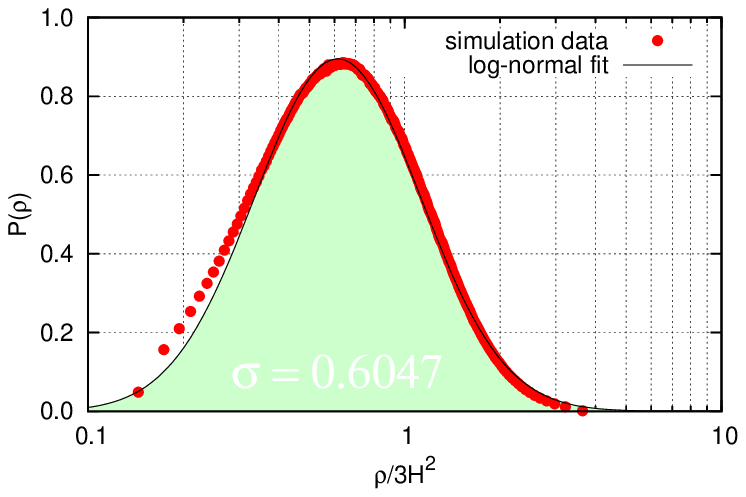, width=6cm} &
    \epsfig{file=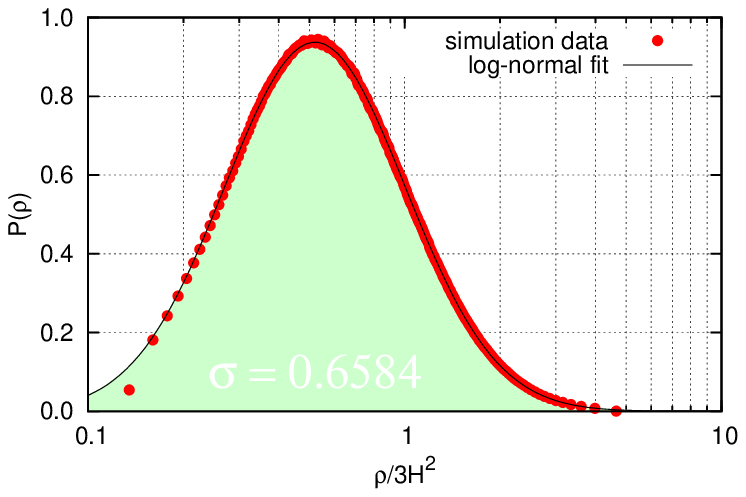, width=6cm} \\
    ~~(c)\hfill $V = \frac{1}{4}\, \lambda (\phi^2 + \chi^2)^2$ &
    ~~(d)\hfill $V = \frac{1}{2}\, m^2 \phi^2 + \frac{1}{2}\, \sigma \phi \chi^2 + \frac{1}{4}\, \lambda \chi^4$ \\
    \epsfig{file=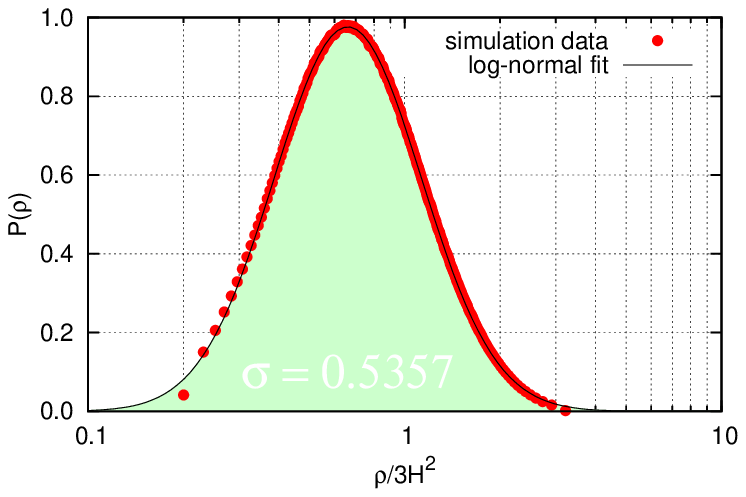, width=6cm} &
    \epsfig{file=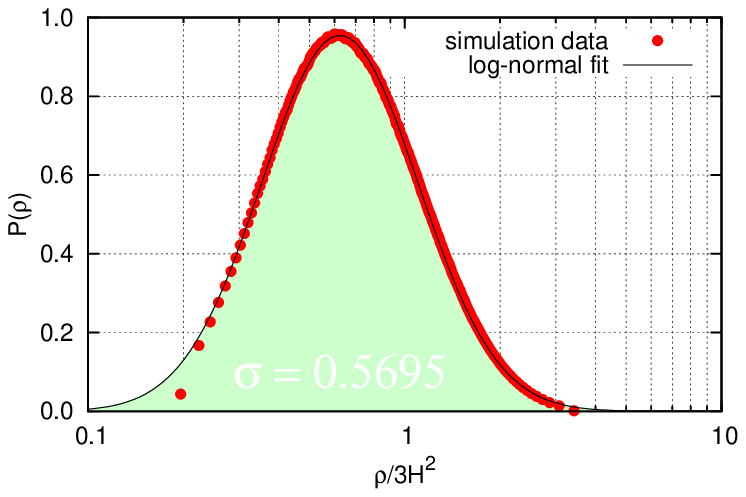, width=6cm} \\
  \end{tabular}
  \end{center}
  \caption{Universality of lognormal density distribution in various two-field preheating models with inflaton decaying via broad parametric resonance.}
  \label{fig:universality}
\end{figure}

While dynamics of the non-equilibrium phase transition can be rather varied,
late stages of preheating appear to have a certain universality to them. After
initial transient, the field evolution leads to a highly inhomogeneous state
similar to that shown in Figures~\ref{fig:rho}b and \ref{fig:strings}b which
persists on long time scales, with slow fragmentation going on. A striking
feature of this regime is that one-point probability distribution function of
energy density contrast $\delta\equiv\rho/\bar{\rho}$ appears to be statistically stationary, and universal
across a class of preheating models \cite{Frolov:2008hy}.
Figure~\ref{fig:universality} shows late-time energy density PDFs (with
dilution due to expansion scaled out) for four different preheating models
ending via broad parametric resonance. All of them fit lognormal distribution
\begin{equation}\label{eq:lornormal}
  P(\rho)\, d\rho = \frac{1}{\sqrt{2\pi}\,\sigma}\, \exp\left[ - \frac{(\ln\rho - \mu)^2}{2 \sigma^2}\right] \frac{d\rho}{\rho}.
\end{equation}
It is very tempting to attribute scaling and universality of the late stages
of preheating to scalar field turbulence \cite{Micha:2002ey,Micha:2004bv},
especially since lognormal density distributions are known to arise in
relativistic fluid turbulence \cite{Nordlund:1998wj}, but the subject needs
further investigation.

\section{Large-Scale Primordial Fluctuations from Preheating}

\begin{figure}
  \centerline{\epsfig{file=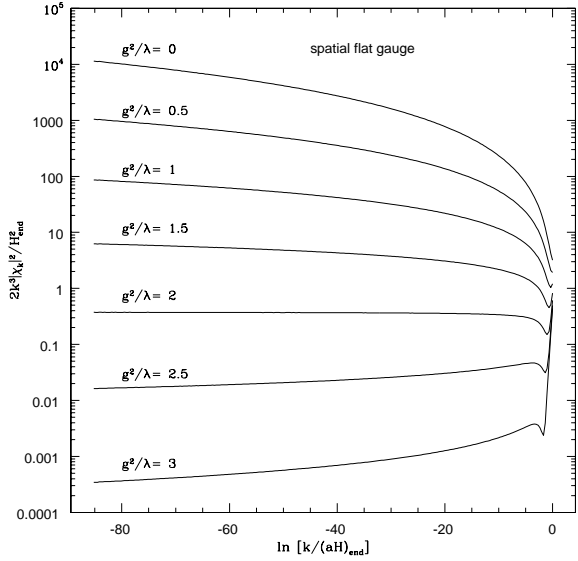, width=8cm}}
  \caption{Primordial fluctuation spectrum of field $\chi$ produced by inflation \cite{Bond:preview}.}
  \label{fig:spectrum}
\end{figure}

As interesting as non-equilibrium field evolution could be, thermalization
wipes out most of the details in the final state after the phase transition.
However, dynamics of the transition could affect the expansion history of the
universe, and leave an imprint in the observable large-scale curvature
fluctuations produced during preheating \cite{Chambers:2007se, Bond:2009xx}.

Inflation transforms sub-horizon quantum vacuum fluctuations in all the light
fields into super-horizon classical fluctuations. These fluctuations are
statistically homogeneous and isotropic Gaussian random fields, and are
completely described by their spectra with amplitude $P(k) \sim H^2/4\pi^2$
evaluated at horizon crossing $H=k/a$. Causally disconnected patches on
super-horizon scales evolve essentially independently, and large-scale
curvature fluctuations $\Phi$ are the difference $\delta N$ in amount of
expansion $N\equiv\ln a$ different patches experience from the constant
curvature hypersurface at the end of inflation to the constant density (and
temperature) hypersurface once thermalization occurred
\cite{Starobinsky:1982ee, Salopek:1990jq, Sasaki:1995aw}. Fluctuations of the
inflaton $\delta\phi$ are usually the main source of metric curvature
fluctuations $\Phi$, with their amplitude enhanced by the slow-roll parameter
$P_\Phi(k) = P_\phi(k)/(2m_{\text{pl}}^2\epsilon)$. It is also entirely
possible to convert isocurvature modes from subdominant light fields into
observable curvature perturbations, for example as it happens in curvaton-type
scenarios \cite{Linde:1996gt, Lyth:2001nq} and modulated reheating
\cite{Dvali:2003em, Kofman:2003nx}. Resonant preheating dynamics can create
and significantly amplify curvature fluctuations from isocurvature modes of
light fields, as suggested by \cite{Chambers:2007se, Chambers:2008gu} and
calculated in \cite{Bond:2009xx, Bond:preview}.

For simple preheating model (\ref{eq:V:L4G22}) with small values of coupling
$g^2/\lambda$, the second field $\chi$ is light during inflation, and acquires
fluctuation spectrum with power on super-horizon scales comparable to
inflaton, as shown in Figure~\ref{fig:spectrum}. Super-horizon fluctuations of
$\chi$ are converted to curvature fluctuations through preheating dynamics.
The basic mechanism is that the flat $\chi$-direction of potential
(\ref{eq:V:L4G22}) is suddenly lifted due to expectation value of
inhomogeneous terms like $\langle\delta\phi^2\rangle$ when parametric
resonance instability develops \cite{Bond:2009xx}. This will modulate equation
of state based on the value of homogeneous mode in field $\chi$ at a time
inhomogeneity develops, and create curvature fluctuations dependent on initial
value of $\chi$ on super-horizon scales.

\begin{figure}
  \centerline{\epsfig{file=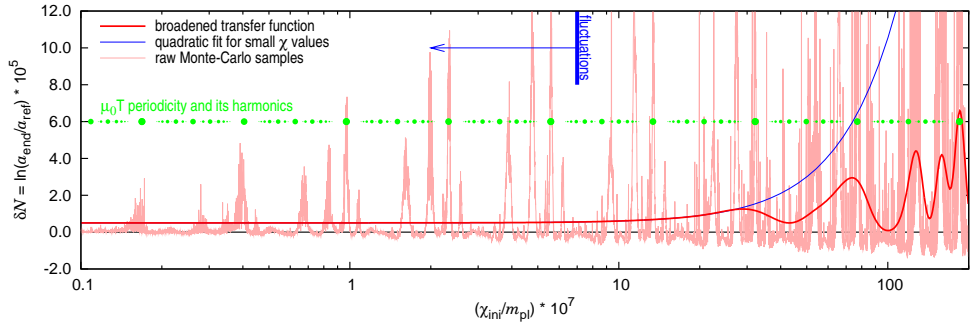, width=\textwidth}}
  \caption{Non-linear transfer function $F_{\text{NL}}(\chi)$ connecting initial value of super-horizon mode $\chi_{\text{ini}}$ with curvature fluctuation $\delta N$ it produces \cite{Bond:2009xx}. Thick red line shows result of averaging over substructure not resolved in CMB observations.}
  \label{fig:fnl}
\end{figure}

Calculating curvature fluctuations generated by preheating involves tracing
minute differences in expansion history, from the end of inflation to
thermalization, in non-linear three dimensional simulations of regions of the
universe corresponding to different initial values of $\chi$ on super-horizon
scales. This is a very demanding numerical problem, not only in terms of
computing power, but precision required. The first attempt encountered
numerical difficulties \cite{Chambers:2007se, Chambers:2008gu}, and it
required development of new numerical integration techniques to obtain the
answer \cite{Bond:2009xx}. Skipping a lot of technical details which will be
discussed elsewhere \cite{Bond:preview}, the total curvature fluctuation
$\Phi$ produced in the inflation model (\ref{eq:V:L4G22}) is
\begin{equation}\label{eq:Fnl}
  \Phi(\vec{x}) = \Phi_{\text{G}}(\vec{x}) + F_{\text{NL}}\big(\chi_{\text{G}}(\vec{x})\big),
\end{equation}
where $\Phi_{\text{G}}$ is the usual nearly Gaussian contribution from
inflaton fluctuations $\delta\phi$, and the second \textit{uncorrelated} term
is generated by preheating from the super-horizon mode of the field $\chi$.
The exact distribution of the field $\chi$ sampled on observable part of the
sky depends on inflation history in extreme version of cosmic variance.
The transfer function $F_{\text{NL}}$ shown in Figure~\ref{fig:fnl} is quite
non-linear and could lead to non-Gaussian fluctuations of
the form very different from the usual weak non-Gaussianity parametrization
\begin{equation}\label{eq:fnl}
  \Phi(\vec{x}) = \Phi_{\text{G}}(\vec{x}) + f_{\text{NL}}\Phi_{\text{G}}^2(\vec{x}).
\end{equation}
The amplitude of curvature fluctuations produced by preheating in model
(\ref{eq:V:L4G22}) is $10^{-5}$, which is comparable to the curvature
fluctuations from inflaton, so the two could potentially be disentangled
by searching for non-Gaussian component in the observed CMB temperature anisotropy.

\section{Discussion: Going after Observable Signatures of Preheating}

Very little is known about how inflation actually ended, and what was the high
energy physics like at those energy scales (or even what is the inflaton
itself). Traces of reheating are hidden from us by opaque plasma in nearly
thermal state, and are unobservable directly. One must seek signatures of
preheating that survive thermalization and could be detected. These could
include stable relics like topological defects \cite{Kasuya:1998td, Tkachev:1998dc,
Battye:1998xe} or primordial black holes \cite{Rubin:2000dq, Suyama:2004mz, Suyama:2006sr},
stochastic gravitational wave background produced by inhomogeneities during
reheating \cite{Easther:2006gt, Easther:2006vd, GarciaBellido:2007dg,
GarciaBellido:2007af, Dufaux:2007pt, Caprini:2007xq}, or anomalies in the
expansion history of the universe imprinted in the primordial curvature
fluctuations \cite{Dvali:2003em, Kofman:2003nx, Chambers:2007se,
Chambers:2008gu, Bond:2009xx, Bond:preview, Kohri:2009ac, Chambers:2009ki}. Of
these, the last effect appears to be the most promising observationally, as
stable relic formation is difficult without spoiling cosmology, and stochastic
gravitational waves are very hard to detect.

Curiously enough, the simple model of preheating (\ref{eq:V:L4G22}) could
generate non-Gaussian curvature fluctuations of observable amplitude which are
intermittent, producing primordial ``cold spots'' \cite{Bond:2009xx}. Although
arguable \cite{Bennett:2010jb}, the CMB temperature map appears to have slight
statistical anomalies in the form of a cold spot in the southern hemisphere
\cite{Cruz:2009nd}, and slight discrepancy in north-south temperature
anisotropy spectra \cite{Eriksen:2007pc}. Exciting possibility is that
primordial ``cold spot'', whether from preheating or some other early universe
source, could potentially explain both. A tell-tale signature of
\textit{primordial} non-Gaussian cold spot is the associated $E$-mode
polarization pattern around it, which might be possible to test with Planck
data \cite{Vielva:2010vn}. Primordial potential ``dips'' would also manifest in formation of large scale structure.

%%% Bibliography

%\newpage
%\nocite{*}
%\bibliographystyle{apsrev}
%\bibliography{global}

\end{document}